    \documentstyle[11pt,paspconf,epsf]{article}

    \def\etal{{\it et al.}}
    
    \def\eg{{\it e.g.}}

    \def\Halpha{H$\alpha$\,}

\begin{document}


    \title{Morphological aspects of star formation in dwarf galaxies}

    \author{Noah Brosch, Ana Heller and Elchanan Almoznino }

    \affil{The Wise Observatory and
    the School of Physics and Astronomy \\ Tel Aviv University, Tel Aviv 69978,
    Israel}


    \begin{abstract}

    We studied the morphology of star formation in dwarf irregular
    galaxies and found that, in general, this takes place on one side
    of a galaxy and far from the center. This is mainly true for low
    surface brightness galaxies; in high surface brightness dwarf
    irregulars the star formation tends to be more centrally
    concentrated, as well as being more intense. We discuss possible
star formation triggers in dwarf irregular galaxies, and evaluate the
    reasons for the peculiar distribution of star forming regions of
    these galaxies. Stochastic star formation, interactions with 
external gas, and tidal interactions appear to be ruled out as responsible
for the asymmetric pattern of star formation. We
 conclude that asymmetry of a dark matter halo
    or of an asymmetric underlying stellar distribution may trigger an
    asymmetric pattern of star formation.

    \end{abstract}

\vspace{3mm}


    \section{Introduction}

    What triggers the star formation in galaxies ? How does it proceed
    once triggered ? Although much has been written on these subjects
    they are by no means much clearer today than they were two or
    three decades ago, when the field of galaxy evolution was just
    beginning to emerge as a branch of astrophysics.

    In principle, the star formation (SF) can be characterized by two
    boundary condition parameters: the initial mass function (IMF) and
    the star formation rate (SFR). The IMF was originally described by
    Salpeter (1955) as a power law. It describes the number of stars
    formed per unit stellar mass and the exponent in the Salpeter
    formulation is --2.35. A number of possible modifications of the
    IMF have been proposed over the years; we shall not review these
    in detail. It suffices to mention modifications by Miller \&
    Scalo (1979), Kennicutt (1983), and Scalo (1986), some of which
    may be represented as piecewise power laws. Parenthetically, we
    also mention the possibility of a metallicity dependence of the
    IMF (\eg \,, Terlevich \& Melnick 1983).

    The SFR has been proposed to be a power law of the gas density
    (Schmidt 1959), but other formulations have also been put up. Some
    other possibilities were a dependence on the gas {\bf surface}
    density, or on this density combined with a dynamical parameter
    (\eg \,, Silk 1997). However, these relations may not always
    represent the most significant dependencies. For instance, in a
    sample of spiral galaxies the SFR, as measured by the strength of
    the \Halpha \, emission, was found to correlate with the blue
    surface brightness (Phillipps \& Disney 1985). A similar
    correlation with starlight for a dwarf galaxy sample was found by
    Hunter \etal \, (1998).

    There have also been numerous suggestions of possible SF triggers. For
    example, Larson (1986) proposed that large scale gravitational
    instabilities, cloud compression by spiral density waves,
    compression by shear forces in a rotating disk, and random cloud
    collisions in the ISM are the major SF triggers. To these one may
    add shock waves from stellar winds and SNe, tidal interactions,
    collisions with other galaxies, ISM stripping, and cooling flow
    accretion in specific galaxies. SF triggering mechanisms were
    reviewed by Elmegreen (1998).

    Because of the proliferation of potential SF triggers, it is
    difficult to understand the process in large, well-established
    galaxies. For this reason we decided to concentrate our efforts in
    studying SF among the dwarf galaxies (DGs). These are devoid of
    differentially-rotating disks, thus the shear forces which act in
    such disks will not be counted among potential trigger mechanisms
    for DGs. In addition,   dwarf irregular galaxies can hardly be 
    classified as ``spirals'', thus an additional potential trigger may
    be ruled out: the density waves.

    By careful selection of the sample galaxies one may probably
    discount tidal triggering of SF if the objects are selected to be
    far from other galaxies. One caveat related to this issue is the
    possibility of delayed star formation, following a soft (distant)
    tidal encounter, if this somehow triggers a rain of gas clouds 
    onto the disk (V\'{a}zquez \& Scalo 1989). Another option is to
    select the sample from a well-defined galaxy environment, where
    tidal triggering should average out among the galaxies of the
    sample. Selection within a cluster of galaxies would
    conform to this requirement but would not yield many dwarf
    irregulars with star formation, as these tend to avoid high galaxy
    density environments.

    We selected our initial sample of galaxies from the Virgo cluster.
    The advantage is that the environment is well-defined, the objects
    are (relatively) nearby, and the Virgo sky region is accessible
    from both north and south hemispheres. The disadvantage is that
    the galaxies are more distant than those in the Local Group, thus
    one does not expect to detect much detail or individual stars. We
    concentrated   in learning about the SF through
    integrated stellar populations from broad-band and \Halpha \, CCD
    imaging.

    Our sample was selected from the Virgo Cluster Catalog (Binggeli
    \etal \, 1985, VCC) with the further proviso that the galaxies
    would have integrated HI non-zero measurements from Hoffman \etal
    \, (1987, 1989) and their heliocentric velocity would be less than
    3,000 km s$^{-1}$. The galaxies form two sub-samples, selected
    according to their surface brightness as reflected by the
    morphological classification in the VCC. The high surface
    brightness (HSB) sub-sample is comprised of objects with the BCD
    classifier (mixed classifications, such as Scd/BCD, are accepted)
    and is described in Almoznino \& Brosch (1998). The low surface
    brightness (LSB) sub-sample is comprised of ImIV and ImV objects
    (mixed classifications, such as dE/ImIV, are accepted) and is
    described in Heller, Almoznino \& Brosch (1998). With an apparent
    magnitude threshold of 17.5 the LSB sub-sample is complete, while
    the BCD sub-sample is representative and contains $\sim$40\% of
    the candidates in the VCC. Our original sample is, therefore,
    representative of the dwarf irregular galaxies (DIGs) in the
    VC.

    \section{Observational data and their interpretation}

    The HSB galaxies were observed at the Wise Observatory (WO) in
    Mizpe-Ramon from 1990 to 1997, with CCD imaging through the B, V,
    R, and I broad bands, and narrow H$\alpha$ bandpasses in the rest
    frame of each galaxy. The LSB galaxies were observed mostly at the WO,
    with a few images obtained at the 6.0-m BTA telescope in Russia.
    The data set used in the analysis described here is derived 
    exclusively from
    \Halpha \, line and off-line continuum images flux-calibrated
    against spectrophotometric standards. The data processing is
    described in detail by Heller \etal \, (1998 and  
    these proceedings).

    We estimate the SFR from the \Halpha \, flux using the formalism
    of Kennicutt \etal \, (1994) for an 18 Mpc distance adopted to
    the VC: SFR=2.93 10$^{11}$ F(H$\alpha$), where F(H$\alpha$) is the
    total line flux in erg s$^{-1}$ cm$^{-2}$ and the SFR is in
    M$_{\odot}$ yr$^{-1}$. Our main finding, explained in more detail
    by Heller \etal \, (below), is that SF takes place both in the HSB
    and LSB samples. The main difference between the two types of
    galaxies is the intensity of the phenomenon; HSB objects have SFRs
    higher by one order of magnitude than LSBs.

    We searched for correlations between the SFR and other observable
    parameters, in order to understand what determines the SFR in
    DIGs. The search is described in detail in Brosch \etal \, (1998a). The
    relevant result is that the most significant correlation of the
    SFR, expressed as the SFR per unit solid angle to compensate for
    a measure of ignorance of the right distance, was not with
    the surface gas density nor with a surface gas density and dynamical 
parameter similar to
    that put forward by Silk (1997), but simply with the blue surface
    brightness. In other words, what seems to regulate the SFR in DIGs
    is the local pre-existing stellar population, as found for large
spiral galaxies by Phillipps \& Disney (1985).

    We confirmed this in Heller \etal \, (1998), where we checked
    specifically for the link between the \Halpha \, flux of an
    individual HII region and the red continuum flux underneath this
    HII region; this showed a very strong correlation, supporting our
    previous result derived with the integrated properties of single
    galaxies. We conclude that  in dwarf irregular galaxies the SFR is
    regulated by the underlying stellar distribution.

    \section{Morphology of star formation}

    A cursory perusal of the net \Halpha \, images collected for our
    entire DIG sample in the VC showed that the HII regions are mostly
     not centrally located on the galaxy image as shown by
    the red continuum image. In many cases the HII regions appear
    right at the edge of a galaxy, mostly to one side.
    In order to quantify this impression and to eliminate possible
    biases, due to specific details of our sample galaxies or of our
    observational procedures, we collected comparison samples of DIGs
    with \Halpha \, and red continuum intensity distribution
    information from the literature and analyzed them in exactly the
    same manner as used for the VC DIGs. The comparison samples and
    their analysis are described in Brosch \etal \, (1998b).

    The quantitative analysis required the definition of two
    morphological indices. The first describes the degree of
    concentration in the distribution of HII regions and the second
    represents their measure of asymmetry. In both cases the
    reference is the red continuum intensity distribution in the
    galaxy image and the indices are derived using number counts
    of HII regions.
    We first determined the center and approximate extent of the red
    continuum image of each galaxy. This was done by eye-fitting an
    ellipse to the outermost visible contour of the galaxy's red
    continuum image and transposing this to the net-\Halpha \, image.
    We then counted the number of HII regions in the inner part of a
    galaxy, which we defined as the ellipse with the same center and
    axial ratio as the outer contour ellipse but with axes half the
    size of those of the outer contour ellipse. The number of HII
    regions in the inner part of the galaxy was divided by one-third
    (to compensate for the larger area) of the number of HII regions
    in the outer annulus, the space between the inner ellipse and
    the outer one. This ratio we call CI=concentration index and it can
    range between 0 and $\infty$. A value of CI=0 indicates a galaxy
    with HII regions exclusively in the outer part while CI=$\infty$
    indicates an object with exclusively central \Halpha \, emission.

    A second morphological index, representing the asymmetry in the
    distribution of HII regions, was constructed by counting the
    number of HII regions on two sides of a bisecting line traced
    through the center of the ellipse representing the distribution of
    the red continuum light, which was transposed to the net-\Halpha
    \, image. The position angle of this bisecting line was set so as
    to maximize the ``contrast'' in the number of HII regions between
    the two halves of the image. The asymmetry index AI was then
    obtained as the ratio between the smaller number and the larger.
    AI can range between 0 and 1, with a zero value indicating an
    extremely asymmetric distribution having all HII regions to one
    side of the bisector line and a value of 1 representing a fully
    symmetric number distribution.

    The distribution of galaxies in the AI-CI plane is shown in Figure
    1. It is clear that most objects concentrate at low-AI and
    low-CI values, indicating that the HII regions are arranged near
    the edge of a DIG and mostly to one side of it. The points at very 
    low CI are those with CI=0, i.e., galaxies with HII regions only 
in their outer parts. The concentration at (CI=100, AI=0) represents points 
    with exclusively nuclear \Halpha \, emission; these are essentially 
    BCDs and have CI=$\infty$.


\begin{figure} 
\vspace{12cm}
\includegraphics{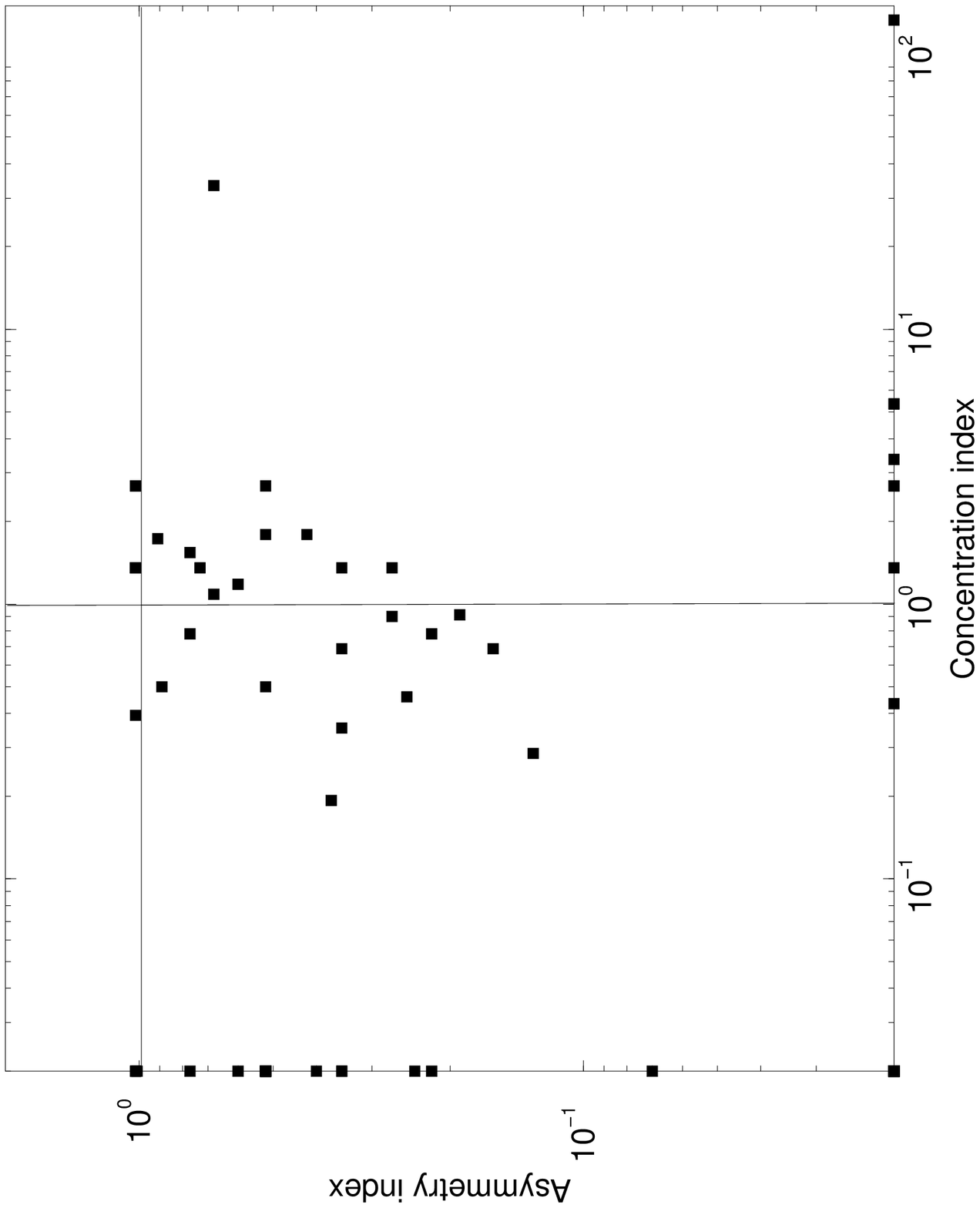}
\caption{Distribution of the asymmetry and concentration indices. We indicate AI=1 (i.e., a
fully symmetric distribution of HII regions) by a horizontal line. The vertical line
indicates CI=1 (i.e., an equal number density of HII regions between the inner and outer
parts of a galaxy). Most of the objects 
concentrate in the low AI bins, indicating a preference 
for asymmetric distribution of HII regions.}
\end{figure}


The tendency of DIGs to have asymmetric distributions of HII regions
is emphasized by Figure 2, which is a histogram of the distribution of 
the AI index. Far 
from being a symmetric Gaussian around AI=0.5, as one could expect for a random
distribution of spotty HII regions over the galaxies, the figure shows most
objects with AI$\leq$0.5. If we eliminate the bin with AI$\approx$0, which 
represents those objects with $\sim$one HII region, we have slightly
more galaxies in the AI$\geq$0.5 part of the distribution but the difference is
not significant; the strong indication of
assymetry is thus driven by  objects with few or single HII regions 
where the SF is  generally not centered on the red light distribution. 
Objects which are not ``extreme'', i.e., AI$\neq$0 and CI$\neq\infty$,
seem to define a tendency of a more symmetric SF pattern  the more 
centrally concentrated the SF is.


\begin{figure} 
\vspace{12cm}
\includegraphics{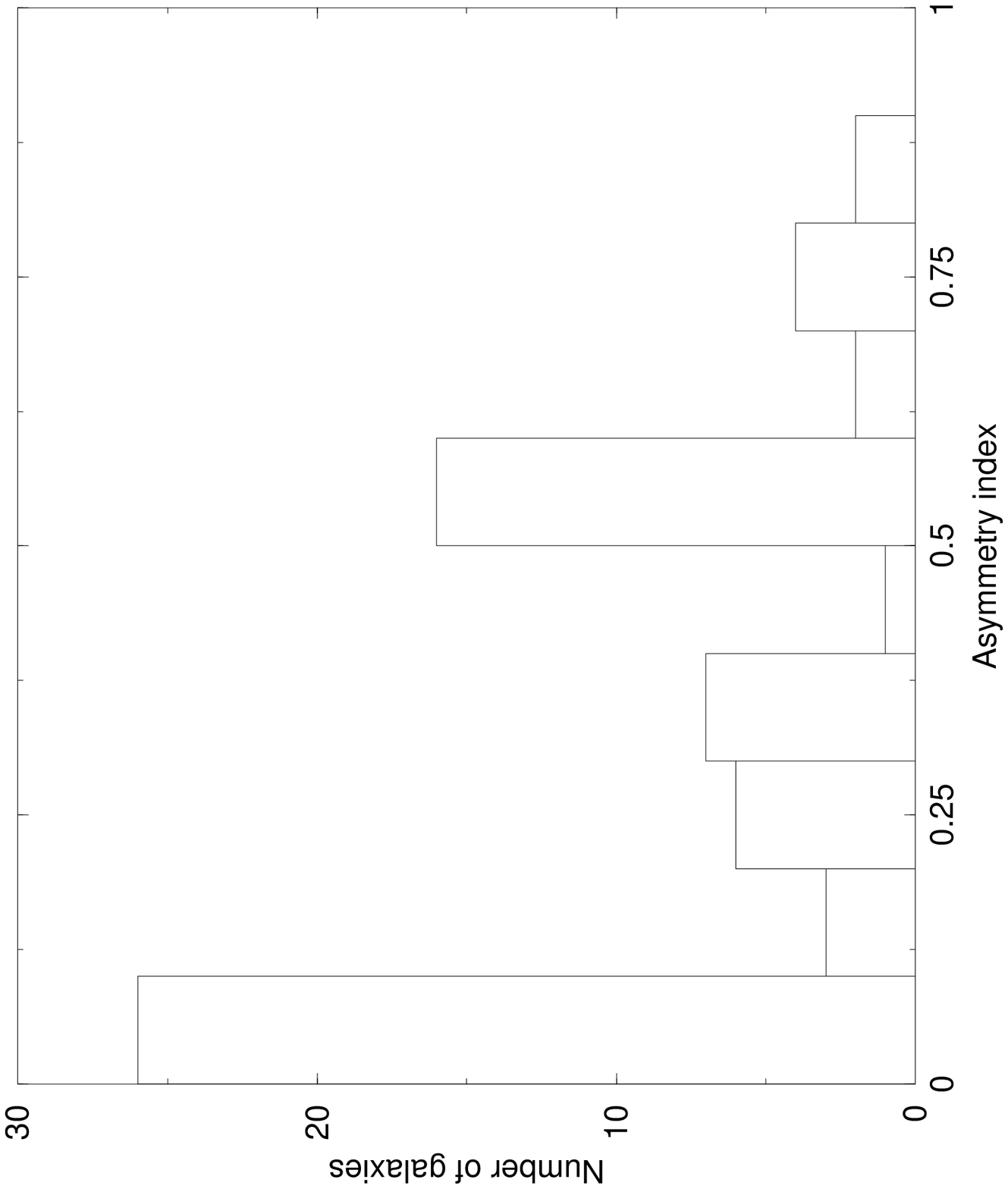}
\caption{Histogram of the asymmetry index for all the sample galaxies, indicating that the
majority have AI$\leq$0.5. This behavior is driven by the large number of objects with a few, 
or just one, HII region. In such cases, the HII regions have a higher probability of not being
located exactly in a symmetric location with respect to the red light distribution.}
\end{figure}

 
    \section{Discussion} 

    We have demonstrated that there is inherent asymmetry in the
    distribution of star-forming regions in DIGs, and that these
    regions tend to reside in the outer parts of these galaxies. This
    is {\bf not} a new discovery, as Hodge (1969) already remarked on
    the asymmetrical distribution of HII in a small number of DIGs.
    His seven objects were selected to be very near and could be analyzed on
    well-resolved photographic images. He checked the asymmetry by
    counting HII regions on different sides of a galaxy, where the
    reference mid-galaxy bisector was that of the HII region
    distribution itself. However, Hodge did not discuss the possible
    origin of this asymmetry.

    The asymmetry and degree of concentration of light in galaxies
    have been used as morphological indicators for the barely resolved
    galaxies of the HST Medium Deep Survey (MDS: Abraham \etal \,
    1996). There are differences between these indices and those used
    by us: firstly, the analyzed MDS images are broad-band I (representing the
    distribution of evolved stars), while we
    use net-\Halpha \, images (i.e., the pattern of newly formed stars). 
    Secondly, our indices are based on eyeball number
    counts of HII regions, while those of Abraham \etal \, are
    calculated by ``impartial'' algorithms from fluxed images. The
    comparison sample at low redshift, from which Abraham \etal \,
    derived their calibration of the A \& C indices against
    morphological types, lacks a good representation of irregular
    galaxies. Nevertheles, they concluded that the 
    late-type galaxies (Sdm and Irr) in the MDS are concentrated at low A \& C
    values, just as we find for our star-forming DIGs. This, they
    proposed, is evidence for the evolution of irregular galaxies.

Gerola \& Seiden (1978) proposed the stochastic self-propagating
SF (SSPSF) as a possible regulatory mechanism. Their simulations, as well as those
by Jungwiert \& Palous (1994), produce preferentially flocculent or
grand-design spirals. There are no specific calculations for
DIGs which show  ``snapshots'' of the SF proceeding with time through 
the galaxy. If the galaxy is small, 
and a number of SNs explode off its center, it may be possible 
for a compression wave to form stars in
suitable location while escaping from the galaxy in places where the
ISM is thin or altogether absent.  This could, in principle, give rise 
to an asymmetric pattern of SF and can be examined
in a nearby DIG (Ho II: Puche \etal \, 1992). The H$\alpha$ and off-line
images show that Ho II forms stars near its center. The H$\alpha$ emission originates
either at the interfaces between large holes in the HI distribution
or in small HI holes. Thus this case does not support 
the SSPSF forming stars asymmetrically or at the edges of a galaxy.

      O'Neil, Bothun \& Schombert (1998) argued that
    the SF trigger in LSB disks is distant tidal interactions.
    Specifically, they combined arguments from V\'{a}zquez \& Scalo
    (1989), that tidally-induced starbursts may be considerably
    delayed after a gas density enhancement, with simulations from
    Mihos, de Block \& McGaugh (1997) who showed that collisions of
    LSBs with compact galaxies produce long-lived disturbances in the
    LSB disks. Although this may operate in large LSB galaxies,
    serving there as SF triggers, the mechanism is probably not
    relevant for SF in DIGs. This is because disky DIGs tend to be
    more stable than large galaxies. Also, the
    simulations of Mihos \etal \, (1997) showed that the 
tidally-induced distortions in LSBs form
    mostly spiral arms and inner rings; it is probable that any
subsequent SF event will retain the same geometry which is a form of
    density wave and is not the asymmetry
    which we observe in DIGs.

Icke (1985) proposed a mechanism by which distant tidal interactions may
trigger SF through shocks in the ISM of gas-rich disks. His mechanism is probably not
relevant for the case of DIGs, because the real strength of the interaction is
likely to be much weaker than he estimated. The reason is that the geometrical factor 
{\bf g} which Icke uses to account for the configuration of the interaction is most
likely $<$0.05, and not 0.39 as fitted by Icke. The factor {\bf g} drives the 
strength of the tidal interaction and to keep this at the required level one
needs to correct upward another parameter in the relation. This is probably very
unlikely, making Icke's work not relevant to SF triggers in DIGs. 
These all argue against tidal interactions 
as being SF triggers in DIGs, as Heller et al mention here.

    Rudnick \& Rix (1998) discussed the asymmetry of
    early-type disk galaxies. They used a different algorithm than ours
    or that of Abraham \etal \, (1996) to detect asymmetry; the
    amplitude of the m=1 azimuthal Fourier coefficient of the surface
    brightness in the R-band. Their claim is that the R-band samples
    stellar populations older than 1 Gyr, thus the asymmetry found in
    these disks must be an inherent property of the stellar mass
    distribution. However, this asymmetry in the mass is not followed by them
    in the SF properties.

    A similar form of asymmetry, lopsidednes of a galactic disk, has
    also been studied by Jog (1997). Its cause is the asymmetric
    motion of particles in a lopsided halo. Jog also showed that the
    gravitational coupling of stars and gas would tend to make the gas
    more unstable in such a situation than if the gas should be
    self-gravitating (Jog 1996, 1998). This effect would enhance the
    asymmetry observed in the young stars in comparison with any
    asymmetry shown by the old star component. Another explanation for
    the asymmetry of disks has been proposed by Levine \& Sparke
    (1998). Their scenario has the disk orbiting in an off-center
    location in the galaxy's DM halo and spinnig in a retrograde sense
    to its orbital motion. 
    It is not clear what implication do these models have on the
    asymmetry of SF in DIGs; they refer mostly to collisionless
    particle simulations of disks while the behavior we witness in
    DIGs is manifested by the dissipative component of a galaxy, its
    ISM. However, if the asymmetry of disks or halos implies a similar
asymmetry of the gravitational potential, this could serve as a trigger
for the asymmetric SF we observe in DIGs.

\section{Conclusions}

We have shown that star formation takes place in both HSB and LSB DIGs,
and that the difference between the two flavors of DIGs is the
intensity of the phenomenon. A consideration of a large sample of DIGs
has demonstrated that these galaxies tend to form stars in an
asymmetric pattern at their outer edge. We could not identify the mechanism 
responsible for this behavior, but suggest that some asymmetry in the gravitational
potential of the galaxies may be the cause.

    \section*{Acknowledgements}

    NB is grateful for continued support of the Austrian Friends of
    Tel Aviv University. EA is supported by a special grant from the
    Ministry of Science to develop TAUVEX, a UV imaging experiment.
    AH acknowledges support from the US-Israel Binational Science
    Foundation. Astronomical research at Tel Aviv University
    is partly supported by a grant from the Israel Science Foundation.
Discussions with Federico Ferrini and Simon Pustilnik on the subject of
star formation in DIGs are greatly appreciated.

    \section*{References}

    \begin{description} 

    \item Abraham, R.G., van den Bergh, S., Glazebrook, K., Ellis,
    R.S., Santiago, B.X., Surma, P. \& Griffiths, R.E. 1996, \apjs,
    107, 1.

    \item Almoznino, E. 1996, PhD thesis, Tel Aviv University.

    \item Almoznino, E. \& Brosch, N. 1998, \mnras, 298, 920.

    \item Binggeli,  B., Sandage, A. \& Tammann, G.A. 1985, AJ,  90,
    1681 (VCC).

    \item Brosch, N., Heller, A. \& Almoznino, E. 1998a, \apj, 504, 720.

    \item Brosch, N., Heller, A. \& Almoznino, E. 1998b, \mnras, in press.

\item de Blok, W.J.G., van der Hulst, J.M. \& Bothun, G.D. 1995, \mnras, 274, 235.

    \item Elmegreen, B.G. 1998, in {\it Origins of Galaxies, Stars, Planets
    and Life} (C.E. Woodward, H.A. Thronson, \& M. Shull, eds.), ASP
    series, in press.

\item Gerola, H. \& Seiden, P.E. 1978, \apj, 223, 129.

    \item Heller, A., Almoznino, E. \& Brosch, N. 1998, \mnras, in press.

    \item Hodge, P. 1969, \apj, 156, 847.

    \item  Hoffman, G.L., Helou, G., Salpeter, E.E., Glosson, J. \& Sandage, A.
      1987, \apjs, 63, 247.


    \item Hoffman, G.L., Williams, H.L., Salpeter, E.E., Sandage, A., Binggeli,
     B.  1989, \apjs, 71, 701.

    \item Hunter, D.A., Elmegreen, B.G. \& Baker, A.L. 1998, \apj, 493, 595.

\item Icke, V. 1985, A\&A, 144, 115.

    \item Jog, C.J. 1996, \mnras, 278, 209.

    \item Jog, C.J. 1997, \apj, 488, 642.

    \item Jog, C.J. 1998, private communication.

\item Jungwiert, B. \& Palous, J. 1994, A\&A, 287, 55.

    \item Kennicutt, R.C. 1983, \apj, 272, 54.

    \item Kennicutt, R.C. 1989, \apj, 344, 685.

    \item Kennicutt, R.C., Tamblyn, P. \& Congdon, C.W. 1994, \apj, 435, 22.

    \item Larson, R.B. 1986, \mnras, 218, 409. 

    \item Levine, S.E. \& Sparke, L.S. 1998, \apj, 496, L13.

    \item Loose, H. H. \& Thuan, T.X. 1986 in {\it Star forming dwarf galaxies
    and related objects} (D. Kunth, T.X. Thuan \& J. Tran Thanh Van, eds.),
    Gif sur Yvette: Editions Fronti\`{e}res, p. 73.

    \item Mihos, C., de Blok, W. \& McGaugh, S. 1997, \apj, 477, L79.

    \item Miller, G.E. \& Scalo, J.M. 1979, \apjs, 41, 513.

    \item O'Neil, K., Bothun, G.D. \& Schombert, J. 1998, astro-ph/9808359.

    \item Patterson, R.J. \& Thuan, T.X. 1996, \apjs, 107, 103.

    \item Phillipps, S. \& Disney, M. 1985, \mnras, 217, 435.


    \item Rudnick, G. \& Rix, H.-W. 1998, AJ, 116, 1163.

    \item Salpeter, E.E. 1955, \apj, 121, 161.

    \item Scalo, J.M. 1986, Found. Cosmic Phys., 11, 1.

    \item Schmidt, M. 1959, \apj, 129, 243.

    \item Silk, J. 1997, \apj, 481, 703

    \item Terlevich, R. \& Melnick, J. 1983, ESO preprint no. 264.

    \item V\'{a}zquez, E. \& Scalo, J. 1989, \apj, 343, 644.

    \end{description}
     \end{document}